\documentclass[preprint,pra,showpacs]{revtex4}

\usepackage{amssymb}
\usepackage{amsfonts}
\usepackage{hyperref}
\usepackage{graphicx}
\usepackage{wrapfig}

\begin{document}

\title{Numerical analysis of decoy state quantum key distribution protocols}

\author{Patrick Rice}
\email{prrice@lanl.gov}

\author{Jim Harrington}
\email{jimh@lanl.gov}

\affiliation{Applied Modern Physics (P-21), MS D454, Los Alamos National Laboratory, Los Alamos, NM 87545, USA}

\begin{abstract}
Decoy state protocols are a useful tool for many quantum key distribution systems implemented with weak coherent pulses, allowing significantly better secret bit rates and longer maximum distances.  In this paper we present a method to numerically find optimal three-level protocols, and we examine how the secret bit rate and the optimized parameters are dependent on various system properties, such as session length, transmission loss, and visibility.  Additionally, we show how to modify the decoy state analysis to handle partially distinguishable decoy states as well as uncertainty in the prepared intensities.
\end{abstract}
 

\maketitle

\section{Introduction}

Quantum key distribution (QKD) is a cryptographic protocol that allows two remote parties (Alice and Bob) to generate a random key (a string of bits) so that only Alice and Bob have any information about the key. The security of the system is guaranteed by the laws of quantum mechanics unlike other key distribution schemes, such as public key cryptography, that rely on assumed computational limitations of the attacker. 
The classic QKD system is BB84 \cite{bb84}, which consists of a sender (Alice) sending a receiver (Bob) a series of bit values encoded on qubits in one of two mutually unbiased bases. They only keep the bit values where Bob measured in the same basis that Alice prepared the state in.  Provided that any eavesdropper (Eve) does not know the basis at the time of transmission, she cannot gain any knowledge of the key Alice sent without introducing a disturbance rate into the stream that Bob measures.  Alice and Bob can then reveal a portion of their bits to determine with high probability (asymptotically $1$) that Eve was not present and thus are assured of unconditional security.  Of course, in non-ideal (real world) situations, noise has to be accounted for, whether it is due to Eve or the physical apparatus. It has been shown \cite{shor-preskill, gllp, ilm} that Alice and Bob can still be assured of unconditional security in the presence of noise by implementing error correction and privacy amplification by way of classical messages over an authenticated public channel.

There are also forms of attack that are based on the specifics of the physical protocol and how they differ from the ideal BB84 protocol.  One common physical difference is the use of weak coherent light pulses for the encoded bits that Alice sends to Bob, instead of using a qubit space. Since it is currently hard to create a steady deterministic source of single-photon signals (SPS), phase-randomized coherent states with their Poisson distribution of photon number are often used instead.  The photon-number-splitting (PNS) attack takes advantage of an inherent vulnerability in these types of signals, specifically the fact that sometimes multiple photons are simultaneously prepared and sent with the same encoded information.  In a PNS attack, Eve performs a non-demolition measurement on the photon number of each signal that Alice sends to Bob.  When the number is greater than one, Eve splits a photon off of the signal and stores it in quantum memory.  After Bob has measured the signal and revealed which basis he measured in, Eve can then measure in the same basis and acquire full information on those signals without introducing any disturbance.

In order to protect against this type of attack, Alice and Bob must place a bound on the number of SPS that Bob received.  Then, they must privacy amplify away all other signals.  If Alice prepares all of the signals from a single mean photon number ($\mu$), conservative analysis leads to a very loose bound and hence a significant reduction in secret bit rate and maximum secure distance.  Decoy states were proposed as a solution \cite{hwang03, lo05, wang05, harrington05} to boost performance and have since been implemented experimentally \cite{decoyExpA, decoyExpB, decoyExpC, decoyExpD, decoyExpE, decoyExpF, decoyExpG, decoyExpH}. This is where Alice creates multiple signal intensities ($\mu_j$) and randomly varies between these intensities when sending each signal to Bob. By looking at the number of signals of each level that reach Bob, one can characterize a slice of how a channel behaves for a signal intensity $\mu_j$. Since each signal level gives a different slice, the decoy states better characterize the channel. This means that they can place a much tighter bound on the SPS and thus acheive significantly greater rates and maximum distances.

The rate of key generation is affected by several protocol parameters, namely the number of decoy states, the intensity of each decoy state ($\mu_j$), and the proportion of each decoy state among the total signals sent.  It is thus useful to form heuristics for determining their optimal setting.  These optimal parameters are in turn affected by the system parameters: dark (and background) count rate ($y_0$), signal visibility ($V$), the session length in number of signals that Alice sent ($N$), the security parameter ($\epsilon$), and the channel characteristics. By building on the formalism introduced in \cite{harrington05}, this paper will present results from numerical analysis of the effects of system parameters on the optimal decoy state protocols and their associated rates. In Section II, we present our secret key generation rate formula and discuss the decoy state optimization programs. In Section III, we show how we perform the global search over system parameters and find the optimal decoy state scheme. Numerical results detailing how the different system parameters affect the rate are presented in Section IV.  We then demonstrate how to carry out decoy state analysis either when the intensities are not known exactly (Section V) or when the levels of the protocol are partially distinguishable by an adversary (Section VI).  Finally, we summarize the heuristics found in this work in Section VII.  Appendix A contains an argument concerning the validity of the decoy state analysis regardless of whether the quantum channel is stationary or time-varying.

\section{Calculating the rate}

For any given protocol, there is a simple linear program that can be used along with some statistics from Bob's data to place a tight bound on the number of SPS.  There is also a quadratic program to find a bound on the bit error rate (BER) of the SPS.  In both cases we consider the intensities $\mu_j$ to be known exactly; see Section V for a relaxation of this assumption.  The length of the secret key generated is the bound on the SPS plus the dark counts \cite{nothing} minus the amount of information revealed during the error-correction process minus the amount of required privacy amplification.   The following secret key rate formula has a very similar form to that of GLLP \cite{gllp}, but its (composable) security is actually a finite statistics derivation of Koashi's security proof \cite{koashi05,koashi06}, which will be published elsewhere.  For the numerical studies reported in this work we use:
\begin{eqnarray}
K = R N = \sum_{j} \biggl( S_j + D_j   &-& f_{\rm EC} \cdot C_j \cdot  H_2({\rm BER}) 
-
f_{\rm PA} \cdot S_j \cdot H_2(b_{1}^{\text{max}}) \biggr) \,,
 \label{rate}
\end{eqnarray}
where the summation is only of the $j$'s that label signals that encode secret key bits, $K$ is the length of the key, $R$ is the key rate, $N$ is the session length measured in the number of signals sent, $S_j$ is the lower bound on the SPS and $D_j$ is the lower bound on the dark counts, $C_j$ is the number
of total signals that Bob received from decoy state $j$, $H_2(\cdot)$ is the binary Shannon entropy function and $b_1^{\rm max}$ is an upper bound on the bit error rate of only the SPS. Efficiency factors $f_{e.c.}$ and $f_{p.a.}$ respectively relate how close error correction and privacy amplification are to the Shannon limit. The simulations here use $f_{\rm EC} = 1.2$ and $f_{\rm PA}  = 1 + 1.53 (b_{1}^{\text{max}})^{-0.54} S^{-0.44} $, where $S = \sum_j S_j$ is the bound on the total number of SPS contributing to the secret key. (The expression for $f_{\rm PA}$ is a rough numerical fit we found for calculating the number of typical strings needed to describe the output of a binary symmetric channel with high confidence, as needed for computing privacy amplification in Koashi's approach \cite{koashi05}.)
Not every level needs to encode key bits; there can be levels acting purely as decoys, which will contribute by bounding the SPS.  In most of our simulations, only one of the (usually three) levels is chosen to encode information for generating secret key.

Prior to calculating the amount of required privacy amplfication, Alice and Bob will share information on: the number of signals at each level sent by Alice ($N_j: \sum_{j}N_j = N$), the number of signals at each level successfully received by Bob ($C_j$), and the number of errors on the sifted bits at each level ($E_j$). From this data, they must find the lower bound $S$ and the upper bound $b_1^{\rm max}$. 

\subsection{Finding yields}

Since Eve cannot distinguish between the decoy states except by means of the photon number, the channel parameters that she can adjust can be varied independently for each photon number, $k$. Define $y_k$ to be the probability that Bob gets a click at his detector given that the click came from a $k$ signal that Alice sent. $y_0$ will encompass all dark and background counts. If Eve does not apply a stationary channel the $y_k$'s will be averaged probabilities. See Appendix \ref{stationaryArgument} for further details. To bound $S$, first we can bound the probability, $P$, that Bob's click came from a SPS or a dark count given that Alice sent a signal of intensity $\mu_j$. These probabilities are
\begin{eqnarray}
P_j^S &=& e^{-\mu_j} \mu_j y_1 \nonumber \,,\\
P_j^D &=& e^{-\mu_j} y_0\,.
\end{eqnarray}

Since $\mu_j$ is fixed before the protocol begins and Eve is free to pick the values of all the $y_k$'s, the minimum probability, $P_{j}$, is found by ranging over optimization variables $\overline{y}_k$'s to find the minimum subject to the statistics the Bob measured. Each signal, regardless of whether it encodes information, adds a constraint:
\begin{eqnarray}
Y^{-}_{j} \leq e^{-\mu_j} \sum_{k=0}^{\infty} \overline{y}_k \frac{\mu_j^k}{k!} \leq Y^{+}_{j},
\end{eqnarray}
where $Y^{\pm}$ are upper and lower bounds on the overall yield of each decoy state. The maximum likelihood of the overall yield for state $j$ is $\frac{C_j}{N_j}$. The bounds are found such that with probability greater that $(1 - \epsilon)$ the true probability is greater than $Y^-$ and with the same probability it is less than $Y^+$. They can be calculated by using a standard function in statistics, namely the inverse incomplete beta function. 

If there are three decoy states, then when calculating each $P_j$, all three constraints are imposed for each minimization. This is because Eve cannot change the channel based on which decoy state Alice sent so the $y_k$'s must be the same for all decoy states. The variable bounds are $0 \leq \overline{y}_k \leq 1$ since the $y_k$'s are probabilities. The bar indicates a program variable. Putting everything together gives two linear programs:
\begin{eqnarray}
P_j^S &=& e^{-\mu_j} \text{min  } \mu_j \overline{y}_1 \\ 
\text{subject to:} \nonumber \\ 
Y^{-}_{j} &\leq& e^{-\mu_j} \sum_{k=0}^{\infty} \overline{y}_k \frac{\mu_j^k}{k!} ~\leq~ Y^{+}_{j} \ \ \ \forall j \nonumber \\
0 &\leq& \overline{y}_k ~\leq~ 1\ \ \ \forall k \nonumber
\end{eqnarray}
and
\begin{eqnarray}
P_j^D &=& e^{-\mu_j} \text{min  }  \overline{y}_0  \\ 
\text{subject to:} \nonumber \\ 
Y^{-}_{j} &\leq& e^{-\mu_j} \sum_{k=0}^{\infty} \overline{y}_k \frac{\mu_j^k}{k!} ~\leq~ Y^{+}_{j} \ \ \ \forall j \nonumber \\
0 &\leq& \overline{y}_k ~\leq~ 1\ \ \ \forall k. \nonumber
\label{linearEQ}	
\end{eqnarray}
Because it is unwieldy to have $k$ range from $0$ to $\infty$ and thus have an infinite number of optimization variables, we can choose a cutoff value of $K_{{\rm max}}$ and then set any $y_{k \geq K_{{\rm max}}}$ to be either $0$ or $1$ for the upper or lower bounds, respectively. In this paper, $K_{{\rm max}}$ was set to $9$, although numerically there was little difference for $K_{{\rm max}} \geq 5$.

\subsection{Finding error rates}

The next step necessary for calculating the rate is to determine the bit error rate on only the single-photon signals. This can be done in a similar manner. Let $b_k$ be the bit error rate that Eve (or anything else) introduces on signals with $k$ photons present. She can vary her attacks independently for each photon number here as well. The number of errors that Bob measured on signals from state $j$ is $E_j$. Construct bounds, $B^{\pm}$, on the probability that Bob measures an error given that Alice sent a signal. Note that this is not the same as the probability of an error given that Bob received a click. The constraints are
\begin{eqnarray}
B^{-}_{j} \leq e^{-\mu_j} \sum_{k=0}^{\infty} \overline{b}_k \overline{y}_k \frac{\mu_j^k}{k!} \leq B^{+}_{j}.
\end{eqnarray}

Additionally, the constraints on the yields still hold. The total program becomes
\begin{eqnarray}
b_1^{\text{max}} &=& \text{max  }  \overline{b}_1 
\label{bmin}  \\
\text{subject to:}& \nonumber \\
B^{-}_{j} &\leq& e^{-\mu_j} \sum_{k=0}^{\infty} \overline{b}_k \overline{y}_k \frac{\mu_j^k}{k!} ~\leq~ B^{+}_{j} \ \ \ \forall j \nonumber \\
Y^{-}_{j} &\leq& e^{-\mu_j} \sum_{k=0}^{\infty} \overline{y}_k \frac{\mu_j^k}{k!} ~\leq~ Y^{+}_{j} \ \ \ \forall j \nonumber \\
0 &\leq& \overline{y}_k, \ \overline{b}_k ~\leq~ 1\ \ \ \forall k. \nonumber
\end{eqnarray}

This is a quadratically-constrained program, due to the $\overline{y}_k \overline{b}_k$ terms, with a linear objective.  With the double-sided nature of the constraints, this computation takes a bit longer than the simple linear program described above. After calculating $\overline{b}^*_1$, where the $^*$ indicates the optimal value returned from the program, the rate formula (Eq.\ \ref{rate}) can be computed and the length of the secret key and rate of key generation can be determined for any set of measurements generated in a QKD session. Details of this calculation and the protocol used to create this key rate (Eq.\ \ref{rate}) can be found in a forthcoming paper. 

\section{Global search over protocol parameters given system parameters}

In order to find the optimal protocol parameters, a simulation of the particular protocol is done with a set of system parameters. Some of the parameters are relatively straightforward. The security parameter was set at $10^{-7}$ except when testing the effects of its change. The dark count probability $y_0$ included both detector dark counts and background counts. The values used were from $10^{-5}$ to $10^{-8}$. This is the probability of a dark count being detected in the place of one signal. For example, if the combined dark and background rate was $100$ Hz and the system operated at $10$ MHz with no windowing the dark count would be $y_0 = 10^{-5}$. The session length is also expressed in terms of number of signals. The number of signals sent ranged from $10^7$ up to close to $10^{13}$. At a rate of $10$ MHz, the sessions could last anywhere from one second to a few days. Most simulations used a session length of $10^{10}$, which is about $15$ minutes at $10$ MHz. The error rate for dark counts was set at $50\%$. For all other numbers of photons in the signal the bit error rate was a fixed value, ranging from $0\%$ to $5\%$. That is to say that the true values (as opposed to the bounds found in (Eq.\ \ref{bmin})) are $b_0 = .5$ and $b_{k>0} = \frac{1-V}{2}$, where $V$ is the visibility. We also need to define the channel transmission efficiency. The simulations used a simple beamsplitter channel with transmission efficiency $\eta$. This assumption is only for simulating the channel; Bob does not make any assumptions on the channel when performing the decoy state analysis.
 
Once these parameters are set, many simulations are done that range over the protocol parameters. The bulk of the simulations used three signal states; of which one was the vacuum state ($\mu = 0$). The remainder used four states. There are thus four parameters, the signal intensity for the two non-vacuum states and two parameters for the three probabilities of occurrence. Using sample system parameters, many simulations of the rate of key growth were done to search the protocol parameter space. The result was that the rate function appeared convex. The only exception was with a different physical system that will be discussed below; that system had a bimodal rate function. The remaining searches to find optimum protocols used standard convex optimization techniques (the objective function used contains linear and quadratically-constrained programs). 
 
For example, with system parameters: $\epsilon = 10^{-7}$, $N = 10^{10}$, $V$ = $98\%$, $y_0 = 2\times 10^{-6}$, and $\eta = 10^{-3}$, the optimal protocol is found to have the three signal intensity values $(0, 0.063, 0.655)$ occurring with probabilities $(1 \%, 2.75\%, 96.25\%)$. The rate of this protocol is $9.99621 \times 10^{-5}$ which means that the length of the secret key generated is $999,621$. 
 
Some basic features of this example are general for all optimal protocols. The first to note is that one of the signal intensities is $0$. The vacuum state is always an optimal state to use (with three or more levels).  It typically has a small probability less than $5\%$. The next state also has a low probability and a weak signal intensity. The last state has a high probability and a high intensity. This high state does the main work of transmitting the secret key, whereas the lower signal intensities characterize the channel. In fact, such a large share of the key comes from the high signal intensity that any key generated from the weak signal intensity can be ignored (with only slight adjustment in the optimal protocol and secret bit rate).  For the rest of the simulations the high signal state will be assumed to be the only key generating signal.
 
\section{Effects of system parameters on the optimal protocol}

With an algorithm for finding the optimal protocol parameters and the associated rate for a given set of system parameters, the effect of the system parameters on the optimal rate can be examined. 

\begin{figure}[tp]
\includegraphics[width=\columnwidth]{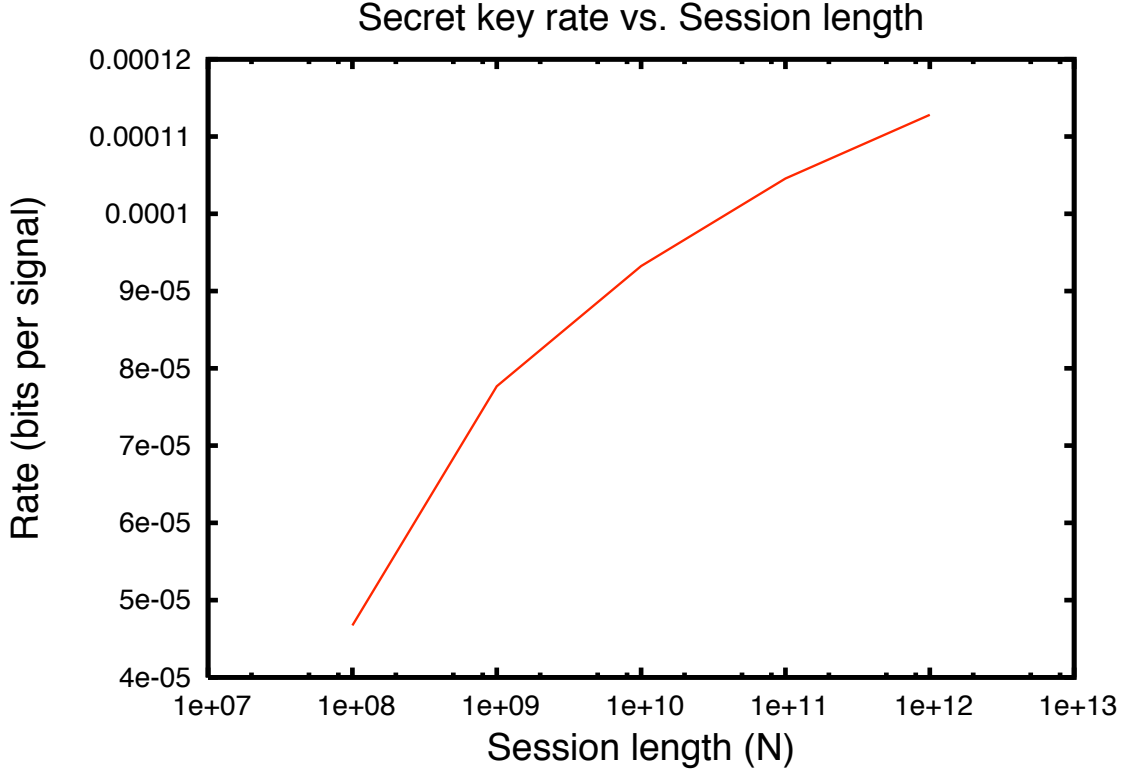}
\caption{(Color online) Semi-log plot of the effects of the session length on the secret key production rate.  The following system parameters are fixed for these simulations: dark count probability $y_0 = 2\times 10^{-6}$, security parameter $\epsilon = 10^{-7}$, visibility $V = 98\%$, and the quantum channel was modeled as a beamsplitter channel with transmission probability $\eta = 10^{-3}$. 
Finite statistic effects (primarily bounds on the single photons and their bit error rate not being tight) lead to a smaller rate for smaller $N$, but these effects become negligible as the session length is increased beyond $10^{12}$, corresponding to nearly a billion received detections.}
\label{graph_N1}
\end{figure}

\begin{figure}[tp]
\includegraphics[width=\columnwidth]{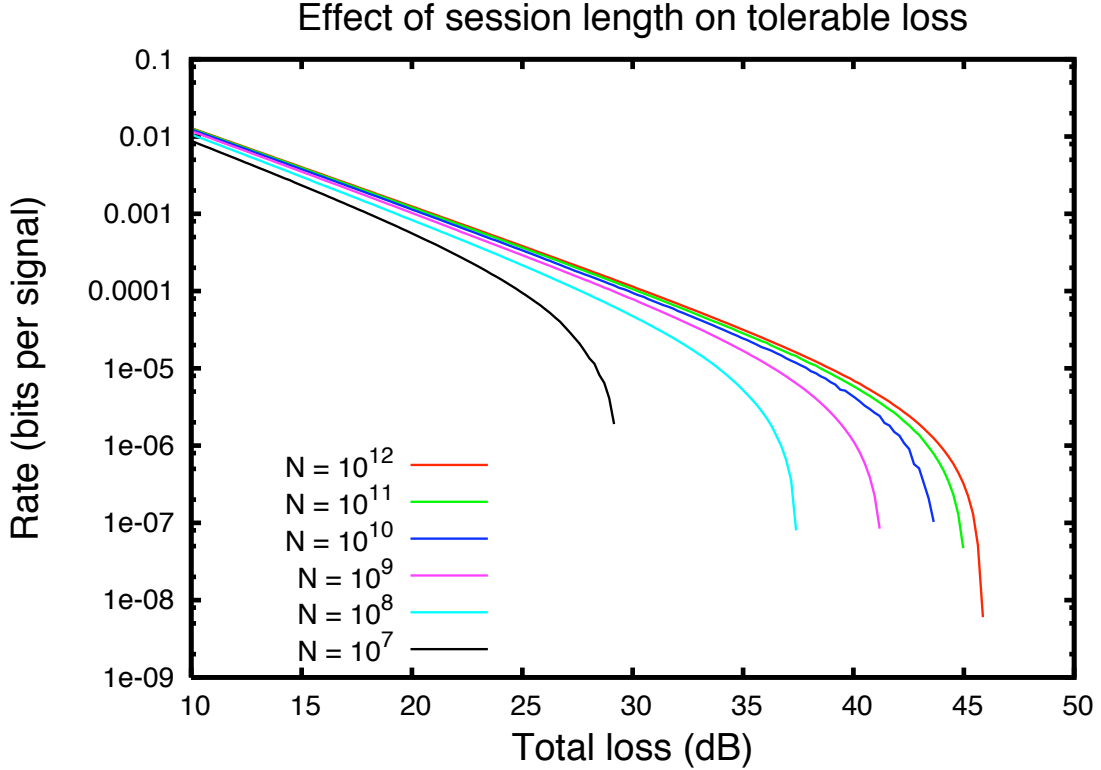}
\caption{(Color online) Semi-log plot of the secret key production rate vs.\ the channel loss for many values of session length $N$ from the best rate with $N = 10^{12}$ down to $N = 10^7$. For short distances (small loss) there is little effect on the rate, but the effect grows much larger at long distances since the amount of count statistics that Bob can gather is on the order of $N\eta$.  For every order of magnitude increase in $N$, the minimum $\eta$ should decrease by an order of magnitude and the maximum channel loss should therefore increase by $10$ dB, but it is quickly damped by other effects.  Fixed system parameters are $y_0 = 2 \times 10^{-6}$, $\epsilon = 10^{-7}$, and $V = 98\%$.}
\label{graph_N2}
\end{figure}

\begin{figure}[tp]
\includegraphics[width=\columnwidth]{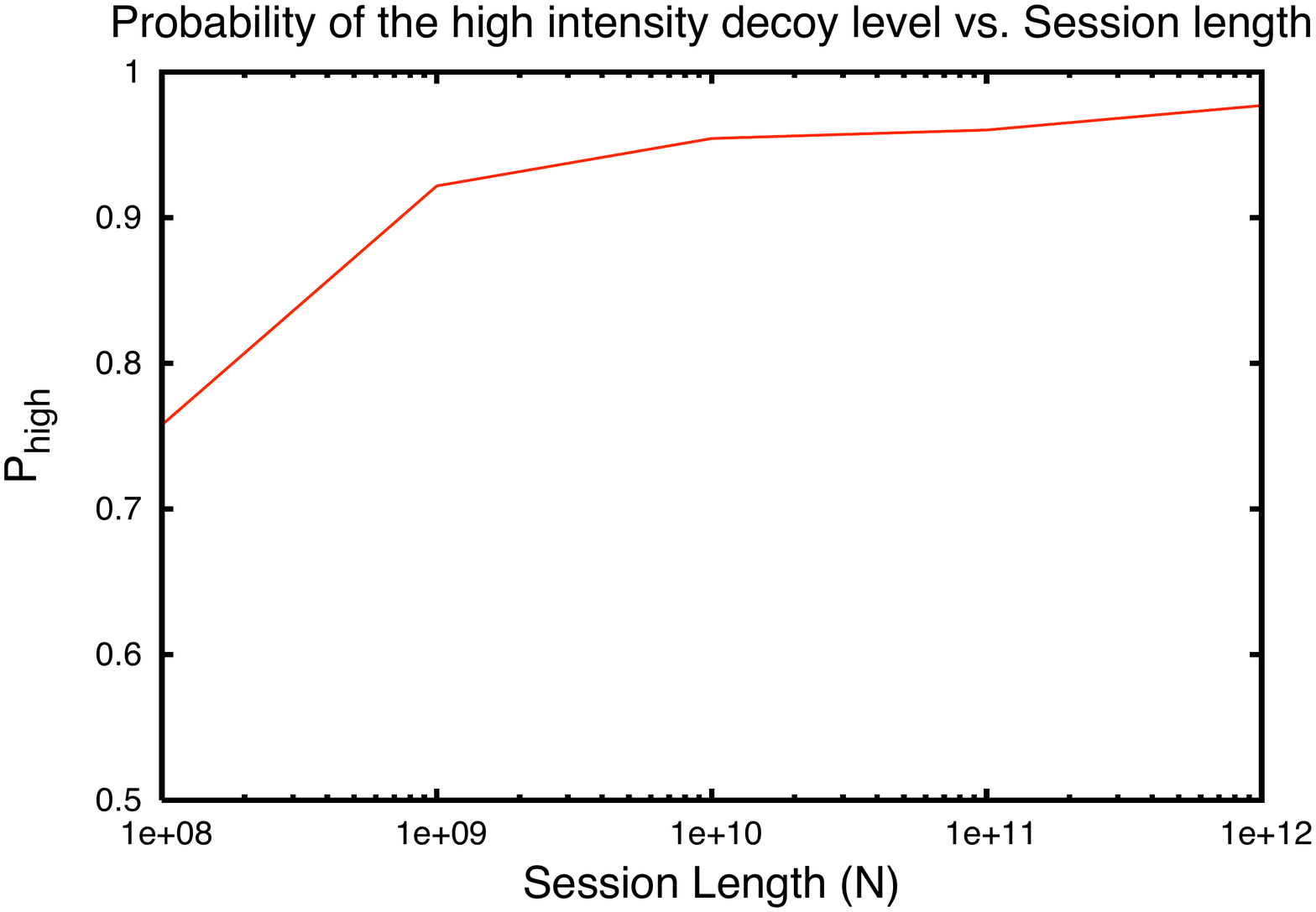}
\caption{(Color online) Semi-log plot of the probability that the high intensity (key-encoding) decoy state. As the number of counts due to the weaker states increase and the bounds get tighter, a higher fraction of the total signals can be used on the high intensity state.  Fixed system parameters are $y_0 = 2\times 10^{-6}$, $\epsilon = 10^{-7}$, $V = 98\%$, and $\eta = 30$ dB.}
\label{graph_N3}
\end{figure}
 
\subsection{Session length}
Increasing the session length effects the rate in two ways. Most clearly it gives Bob additional chances to detect signals and lengthens the raw key and so lengthens the final key. It also increases the rate of final key production per signal sent by Alice. This is due to a tightening of the confidence bounds used in the optimizations. This rate increase effects all protocols and so it shows in an increase of the optimal rate. The main effect on the optimal protocol is to increase the probability of the highest intensity signal when the session length starts out large. The effects of the session length on the rate and probability are displayed in graphs (Figs. \ref{graph_N1}, \ref{graph_N2}).  Generally, the product $N\eta$ needs to at least $10^{5}$ to have enough statistics for successfully producing key.  The increase in the rate and probability of sending the highest intensity signal tends to asymptote around $N \approx 10^{13}$ (Fig.\ \ref{graph_N3}). 
 
\subsection{Security parameter}
\begin{figure}[tp]
\includegraphics[width=\columnwidth]{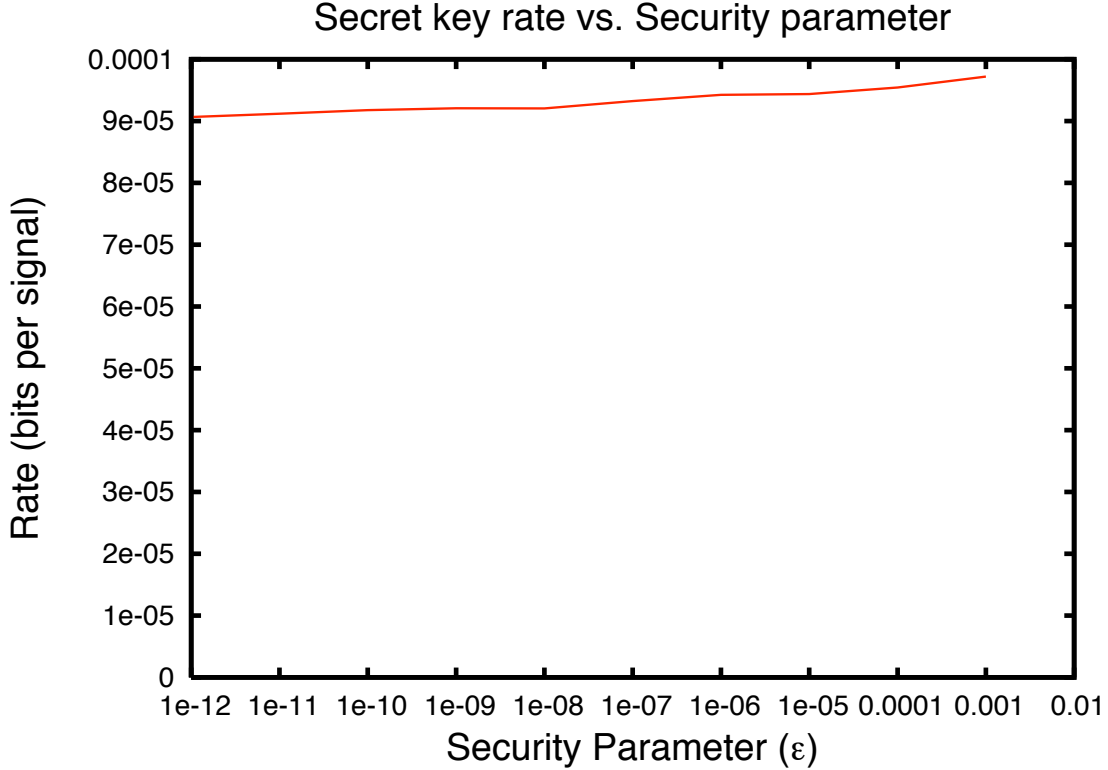}
\caption{(Color online) Semi-log plot of the limited effect of the security parameter on the rate. The rate is quite stable over nine orders of magnitude.  Fixed system parameters are $N=10^{10}$, $y_0 = 2 \times 10^{-6}$, $V = 98\%$, and $\eta = 30$ dB.}
\label{graph_epsilon1}
\end{figure}

 The security parameter effects the rate in a similar manner. As the security parameter is decreased, the confidence bounds tighten and the optimal rate increases.  Fig.\ \ref{graph_epsilon1} shows the small effect of the security parameter on the rate. Perhaps most interesting is the relative stability over nine orders of magnitude.
 
\begin{figure}[tp]
\includegraphics[width=\columnwidth]{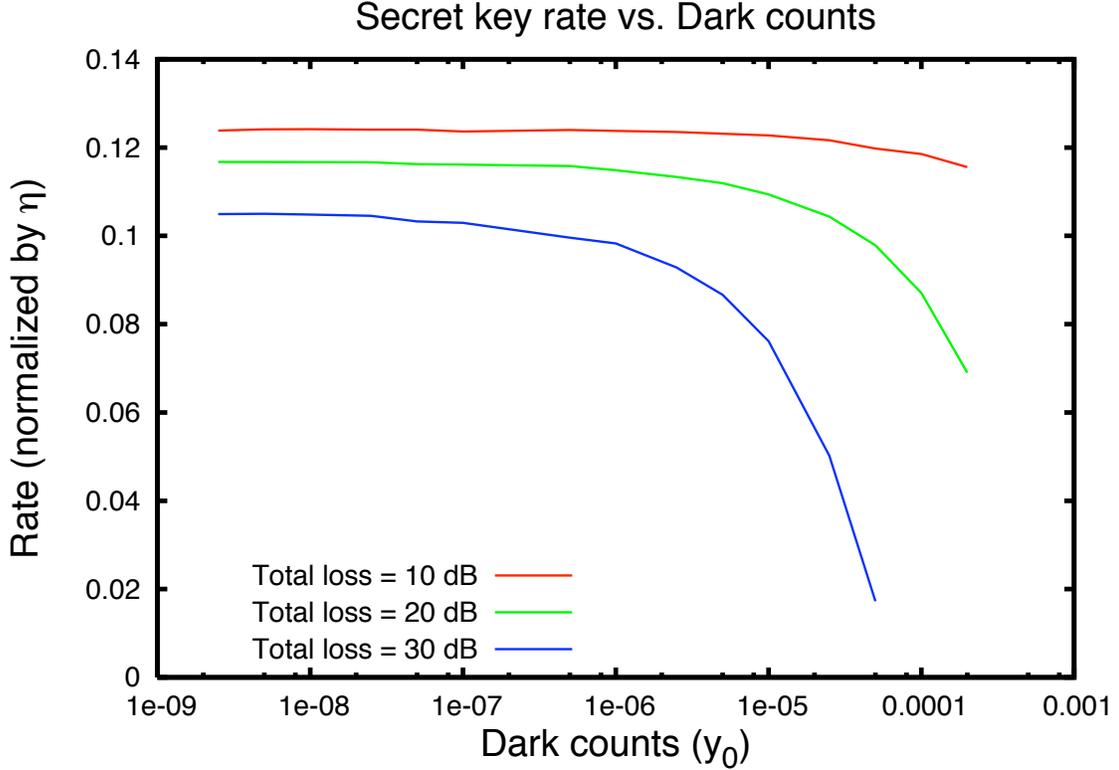}
\caption{(Color online) Semi-log plot of the effect of dark counts on the rate for various amounts of channel loss from low effect at a loss of $10$ dB down to $30$ dB. The rate has been normalized by $\eta$. There is very little negative effect for the small $y_0$ and small channel loss domain, but the effect dominates in the opposite domain.  Fixed system parameters are $N=10^{10}$, $\epsilon = 10^{-7}$, and $V = 98\%$.}
\label{graph_y0-1}
\end{figure}

\begin{figure}[tp]
\includegraphics[width=\columnwidth]{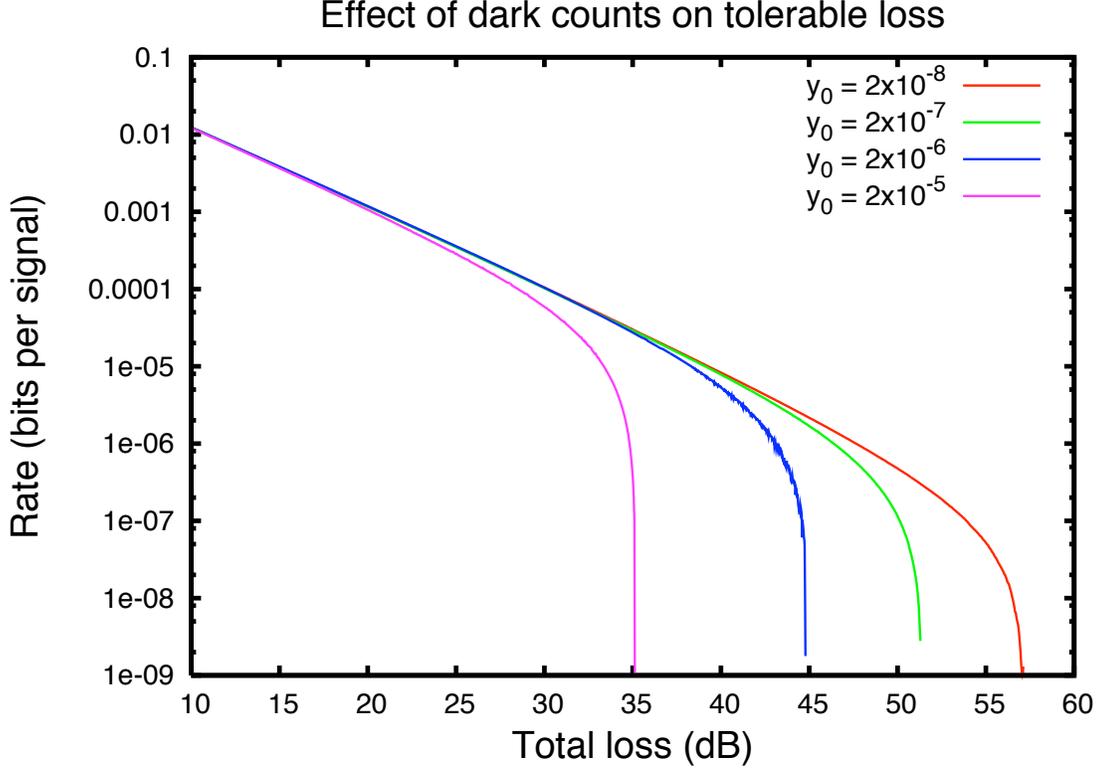}
\caption{(Color online) Semi-log plot of the rate vs.\ the channel loss (dB) for various dark counts from the best performing $y_0 = 2\times 10^{-8}$ down to $y_0 = 2\times 10^{-5}$. The dark count rate severely limits the maximum distance (minimum $\eta$) achievable.  Fixed system parameters are $N=10^{10}$, $\epsilon = 10^{-7}$, and $V =  98\%$.}
\label{graph_y0-2}
\end{figure}

\begin{figure}[tp]
\includegraphics[width=\columnwidth]{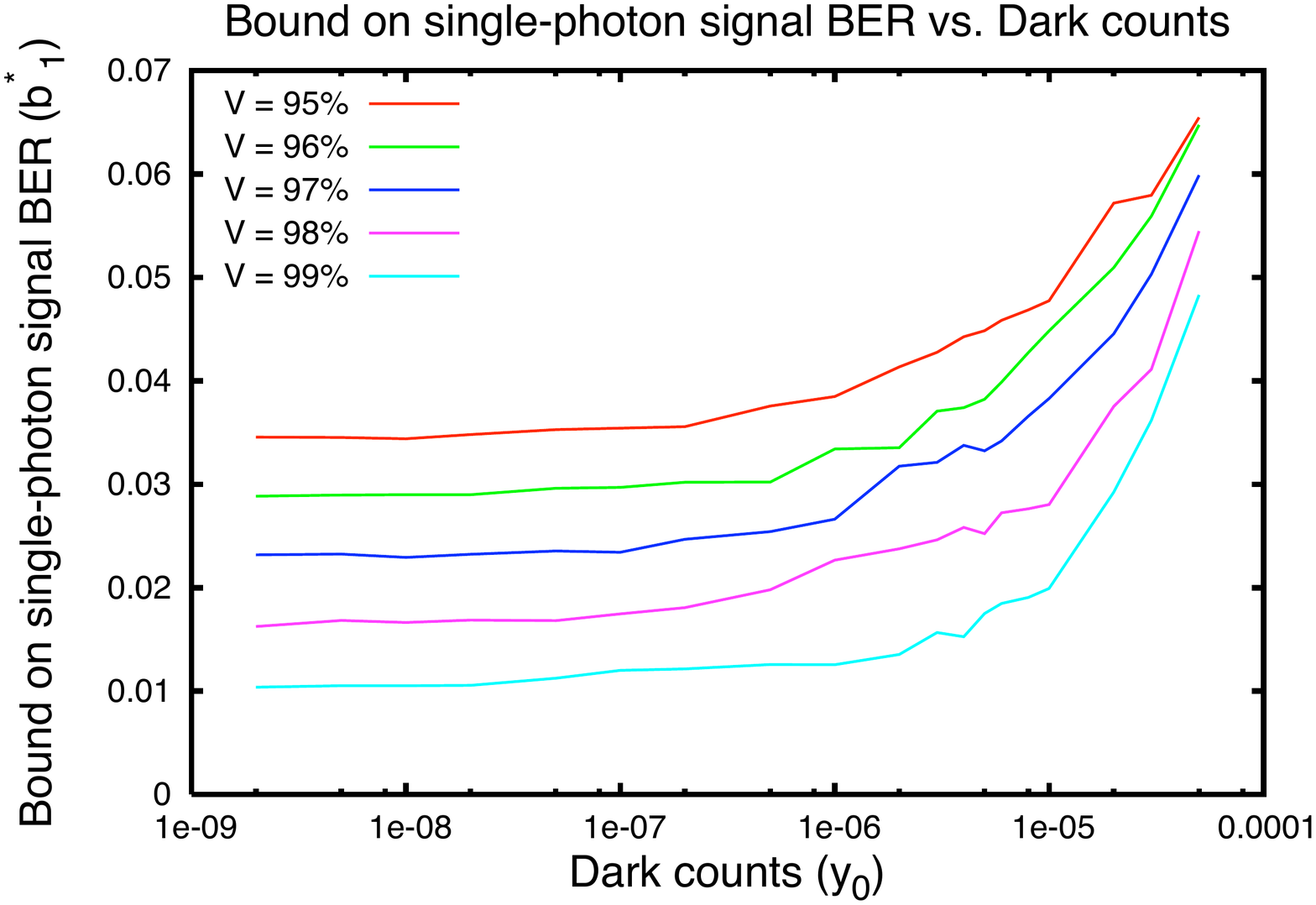}
\caption{(Color online) Semi-log plot of the best bound on the single-photon signal bit error rate ($\overline{b}^*_1$) as a function of the dark counts for visibilities $V = 95\%$ through $V = 99\%$. As the dark count probability becomes a significant fraction of $\eta$, $b_1$ increases dramatically and the secret bit rate drops.  Fixed system parameters are $N=10^{10}$, $\epsilon = 10^{-7}$, and $\eta = 30$ dB.}
\label{graph_y0-3}
\end{figure}

\subsection{Dark Counts}
The dark count rate is the parameter that most determines how far you can conduct QKD. In other words, fixing all other parameters except the transmission coefficient ($\eta$) the value of $y_0$ is the most important parameter to finding the smallest $\eta$ with a positive rate. As the transmission parameter decreases, the number of counts due to signals from Alice decreases but the number due to dark counts does not change. So the proportion of counts due to dark counts increases. Since dark counts have an error rate of $50\%$, this increases the error rate of the raw key significantly. Even with infinite statistics, the secret key rate vanishes when the error rate reaches about $11\%$.  In simulations with reasonable parameters, the smallest $\eta$ was consistently between two and three orders of magnitude larger than the dark count rate. The effect of various dark count rates are shown in graph (Fig.\ \ref{graph_y0-1}, \ref{graph_y0-2}). The dark counts have some effect on the protocol. The lower signal state has to slightly increase intensity so that the dark count noise won't wash out the difference between the low state and the vacuum state. However, most of the effect is indirectly due to increasing the effective error rate (Fig.\ \ref{graph_y0-3}).

\subsection{System visibility}
\begin{figure}[tp]
\includegraphics[width=\columnwidth]{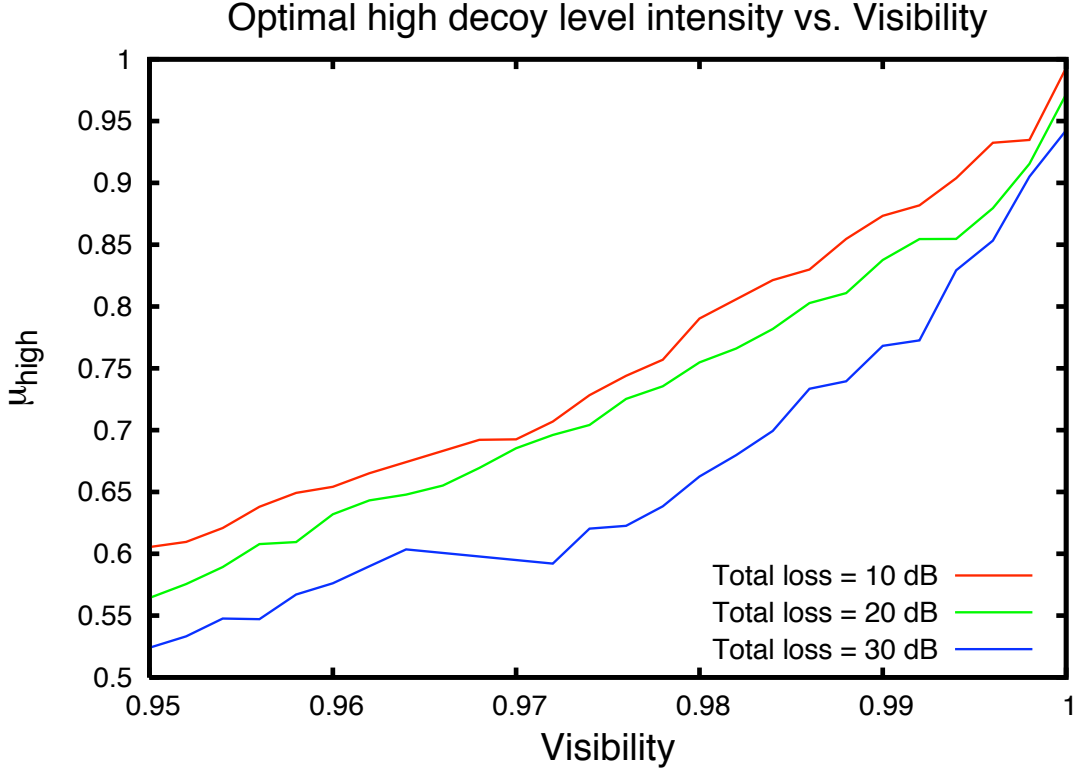}
\caption{(Color online) Plot of the optimal $\mu_{high}$ for various amounts of channel loss from the best performing $10$dB loss down to $30$ dB loss. There is some effect due to the loss, but a much larger effect due to the increase in the visibility can be seen.  Fixed system parameters are $N=10^{10}$, $\epsilon = 10^{-7}$, and $y_0 = 2 \times 10^{-6}$.}
\label{graph_ber2}
\end{figure}

\begin{figure}[tp]
\includegraphics[width=\columnwidth]{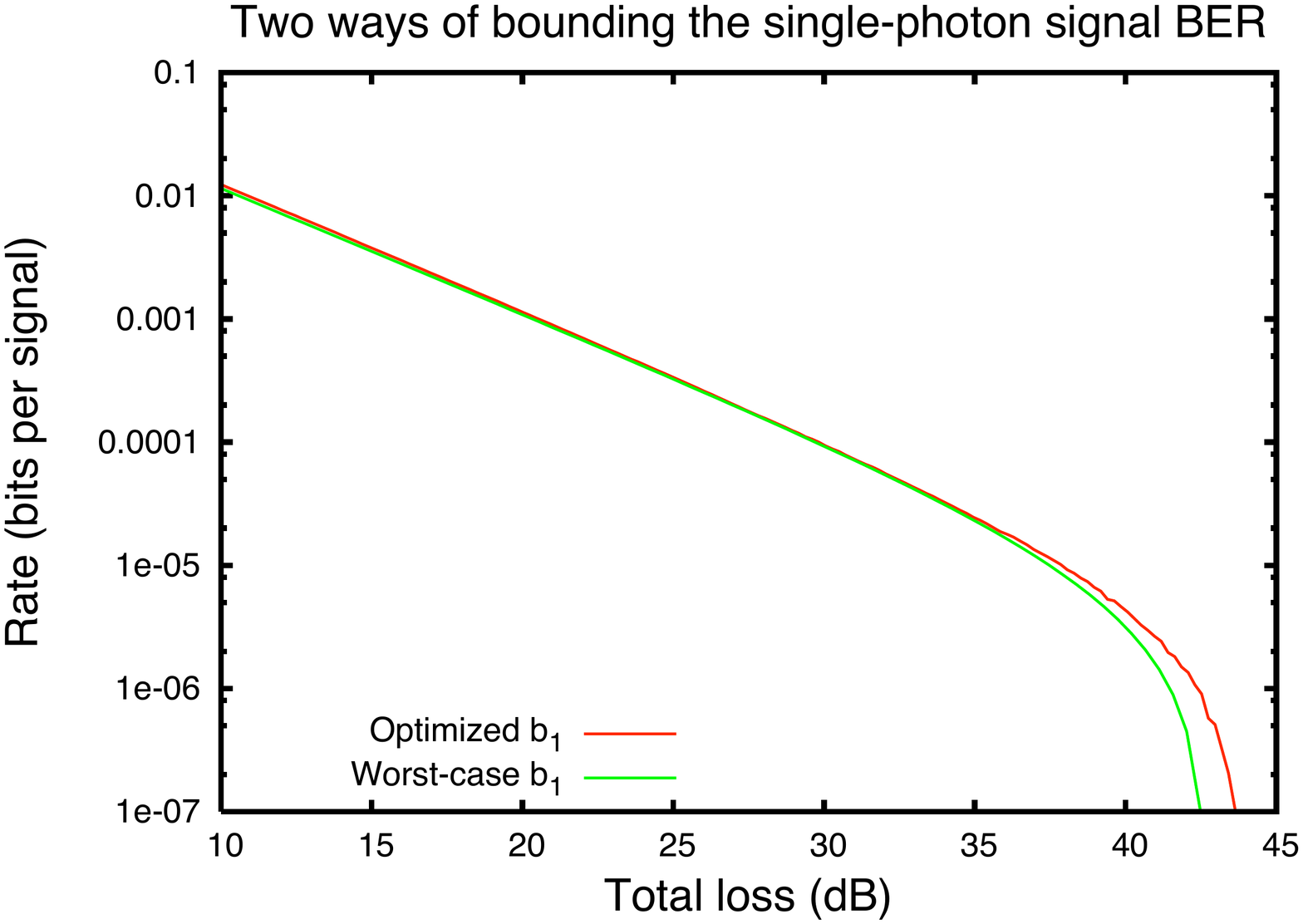}
\caption{(Color online) Semi-log plot comparison between the worst-case bound of the single-photon signal bit error rate and the quadratically-constrained optimization.  Fixed system parameters are $N=10^{10}$, $\epsilon = 10^{-7}$, $y_0 = 2 \times 10^{-6}$, and $V = 98\%$.}
\label{graph_ber3}
\end{figure}

The error rate has the largest effect on which protocol is the optimal protocol, particularly the value of the high state intensity. The secret key rate is also effected by the error rate because the amount of error correction is determined by the error rate (as long as the transmission probability is not within two orders of magnitude as the dark counts). The part of the privacy amplification due to errors is also mainly determined by the error rate. See Fig.\ \ref{graph_ber2} for the optimal rate and the optimal high intensity state. Fitting to the curve gives an approximate formula $\mu_{high} \approx e^{-15(1-V)}$. For larger visibility errors, the ability to bound $b_1$ with the quadratic program is more important. Without the optimization, one must use the worst-case assumption that all the errors observed were errors on SPS. See Fig.\ \ref{graph_ber3} for a comparison of the two methods.

\subsection{Channel transmission efficiency}
\begin{figure}[tp]
\includegraphics[width=\columnwidth]{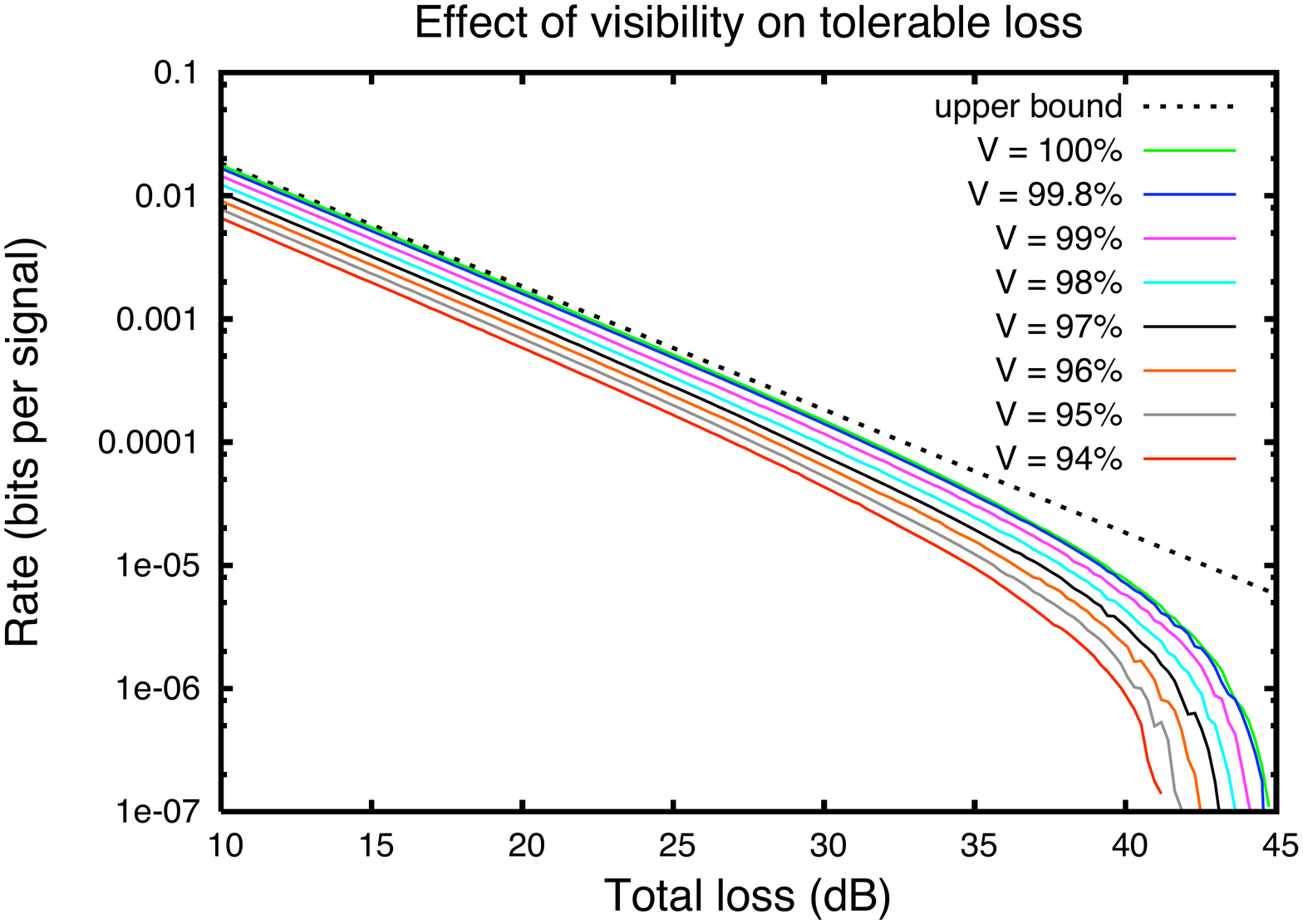}
\caption{(Color online) Semi-log plot of the secret key rate as a function of the channel loss for many visibilities from perfect down to $94\%$. The rate has the most sensitivity to visibility until $\eta$ approaches the dark count rate.  Fixed system parameters are $N=10^{10}$, $\epsilon = 10^{-7}$, and $y_0 = 2 \times 10^{-6}$.}
\label{graph_eta1}
\end{figure}

\begin{figure}[tp]
\includegraphics[width=\columnwidth]{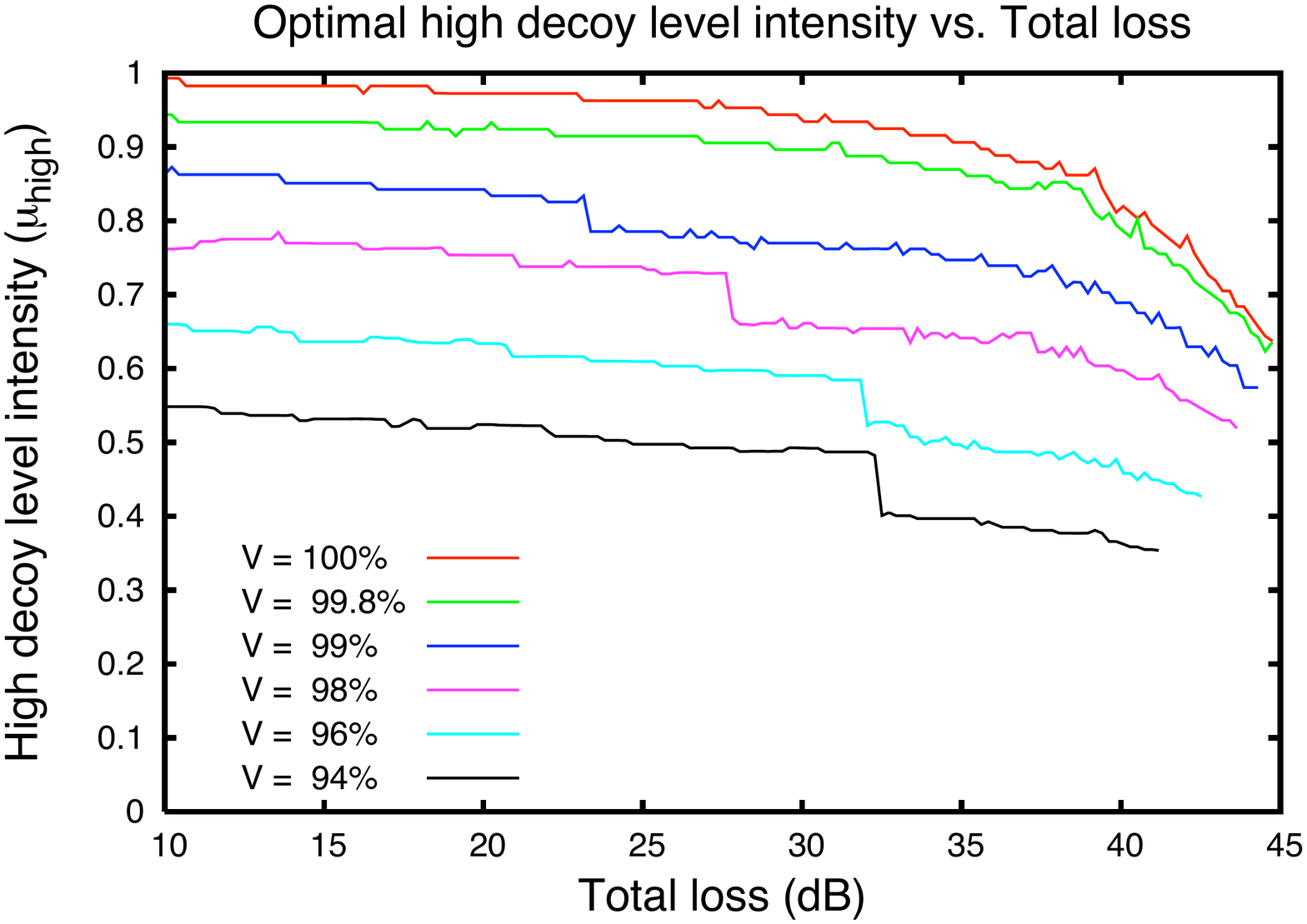}
\caption{(Color online) Plot of the optimal $\mu_{high}$ as a function of the channel loss for many visibilities from perfect down to $94\%$. While the visibility greatly effects the optimal intensity, the level of channel loss does not, except at very high visibility and a high ratio of the dark count probability to $\eta$.  Fixed system parameters are $N=10^{10}$, $\epsilon = 10^{-7}$, and $y_0 = 2 \times 10^{-6}$.}
\label{graph_eta2}
\end{figure}

The transmission coefficient has an effect on the rate but not much on the protocol. See Figs.\ \ref{graph_eta1} and \ref{graph_eta2}. The rate scales linearly with $\eta$, $R = O\left(\eta\right)$. This is the main point in favor of decoy states. Without decoy states, the optimal signal intensity must scale with the transmission efficiency to yield a positive secret bit rate \cite{lutkenhaus}.  With $\mu = O\left(\eta\right)$, then $R = O\left(\eta^2\right)$. With decoy states, the signal intensity (of the high state) shouldn't scale with $\eta$, ($\mu_{high} = O\left(1\right)$). In practice there is still a slight logarithmic dependence on the transmission efficiency. See Fig.\ \ref{graph_eta2}.

\subsection{Detectors}
\begin{figure}[tp]
\includegraphics[width=\columnwidth]{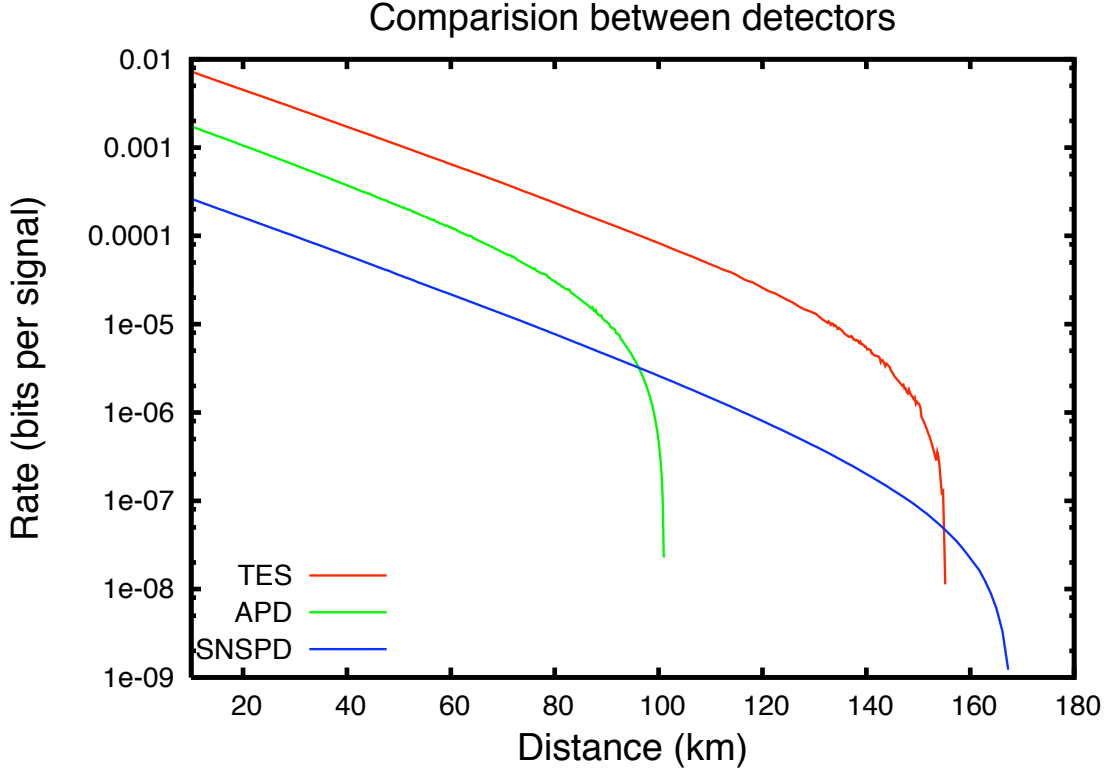}
\caption{(Color online) Semi-log plot of the secret key production rate vs. distance in optical fiber for a variety of different detectors.  The session length is $N = 9 \times 10^{9}$ and the fiber loss is modeled as $0.2$ dB/km.  Additional losses are $7$ dB for the receiving optics and efficiencies of $50\%$,  $10\%$, and $2\%$ for the TES, APD, and SNSPD detectors.  Dark and background count probabilities are $4 \times 10^{-6}$, $1.5 \times 10^{-5}$, and $1.44\times 10^{-8}$, respectively.  Visibility is set at $V = 97.68\%$ and the security parameter is chosen to be $\epsilon = 10^{-7}$.}
\label{graph_detectors}
\end{figure}

We also did simulations detailing the effects of different detectors in a system. The simulations treated the detectors as a plug-n-play portion of the system. The system was assumed to have an optical loss of $7$ dB and a visibility of $97.68\%$, which are measured values for a fiber QKD system developed at LANL. The system was simulated for $15$ minutes at $10$ MHz, even if some detectors are capable of much faster operating rates. The three detectors that were simulated were superconducing nanowire single-photon detectors (SNSPDs), transition-edge sensor superconducting single-photon detectors (TES) and  commercially available InGaAs avalanche photodiodes (APDs). The SNSPDs had a combined dark count probability per signal of $1.44\times 10^{-8}$ and a detector efficiency of $2\%$. The TES detectors had a background count probability per signal of $4 \times 10^{-6}$ (set by blackbody radiation transmitted down the fiber) and a detector efficiency of $50\%$ after applying a filter. The APDs had a combined dark count probability per clock cycle of $1.5 \times 10^{-5}$ and a detector efficiency of $10\%$. The fiber was modeled with a loss of $.2$ dB/km. The simulations are shown in Fig.\ \ref{graph_detectors}. The longest distance acheivable is by the SNSPDs. This is mainly due to their low number of dark counts per signal. The low detector efficiency of SNSPDs however, leads to a lower rate for shorter distances. The APDs could not go out as far as either of the other detectors. The TES detectors have a high detector efficiency and so have a higher rate for most values of channel distance, but they do not reach quite as far as the SNSPDs.

\subsection{Number of levels}
\begin{figure}[tp]
\includegraphics[width=\columnwidth]{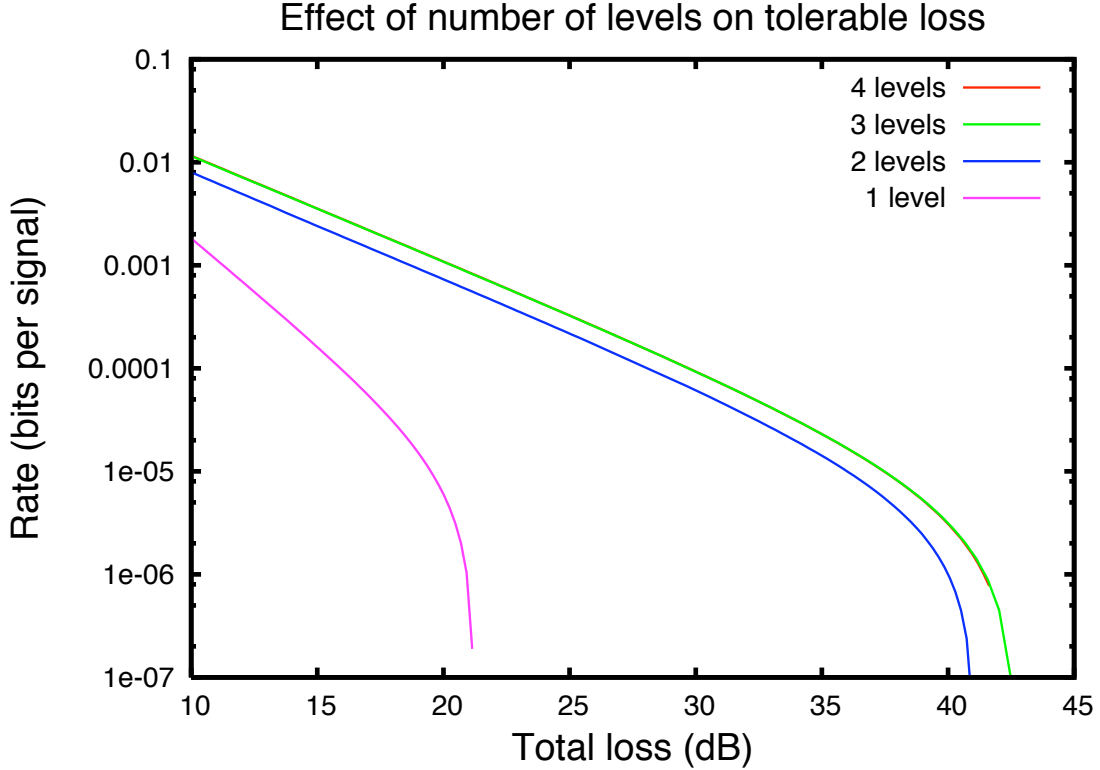}
\caption{(Color online) Semi-log plot of the rate vs.\ the channel loss comparing 1-level (no decoy states), 2-level, 3-level and 4-level protocols.  Adding a weak level alone greatly boosts the performance of the BB84 protocol, and adding a third level (vacuum) raises the rate further.  We have not found any significant difference among optimized 3-level and 4-level protocols, which are overlapping in this plot.  Fixed system parameters are $N=10^{10}$, $\epsilon = 10^{-7}$, $y_0 = 2 \times 10^{-6}$, and $V = 98\%$.}
\label{graph_4-1}
\end{figure}

All results described before were for three-level decoy state protocols. The number of decoy levels that can be implemented is only limited by practical considerations and the fact that increasing the number of levels necessarily decreases the amount of statistics available for each level.  By looking at the effects of the session length, it can be shown that this is not a big concern until the number of states is an extreme case of tens or hundreds. The rate increase of using four states instead of three states is shown in Fig.\ \ref{graph_4-1} and is practically nonexistent. The improvement is small with less than a percent difference. The form of the optimal four-level protocol is similar to the three-level one as well.  Like the previous protocols, there is a vacuum state and a state with low intensity, both with low probabilities. The three-level protocol's high intensity state splits into two high states in the four-level protocol. The two high states have intensities that straddle that of the three-level protocol (one intensity is higher and one is slightly lower).  Generally, the highest state gets most of the probability. 

For both the three- and the four- level protocols, there was an advantage to only sending information with the high intensity levels and allowing the low intensity and vacuum levels to be pure decoys. This was not the case with the other two results shown on Fig.\ \ref{graph_4-1}. With only one level, that level must send information, and in the two-level protocol both levels are sending information. As shown in the figure, adding even one decoy level dramatically protects against the photon-number splitting attacks and significantly improves the maximum tolerable loss. Adding another level yields the three-level protocol that is the focus of this paper and an improvement over the two-level protocol. After that, it seems that the asymptotic limit is reached very quickly with very little noticeable improvement with a four-level protocol.

\section{Uncertainty in intensities}

\begin{figure}[tp]
\includegraphics[width=\columnwidth]{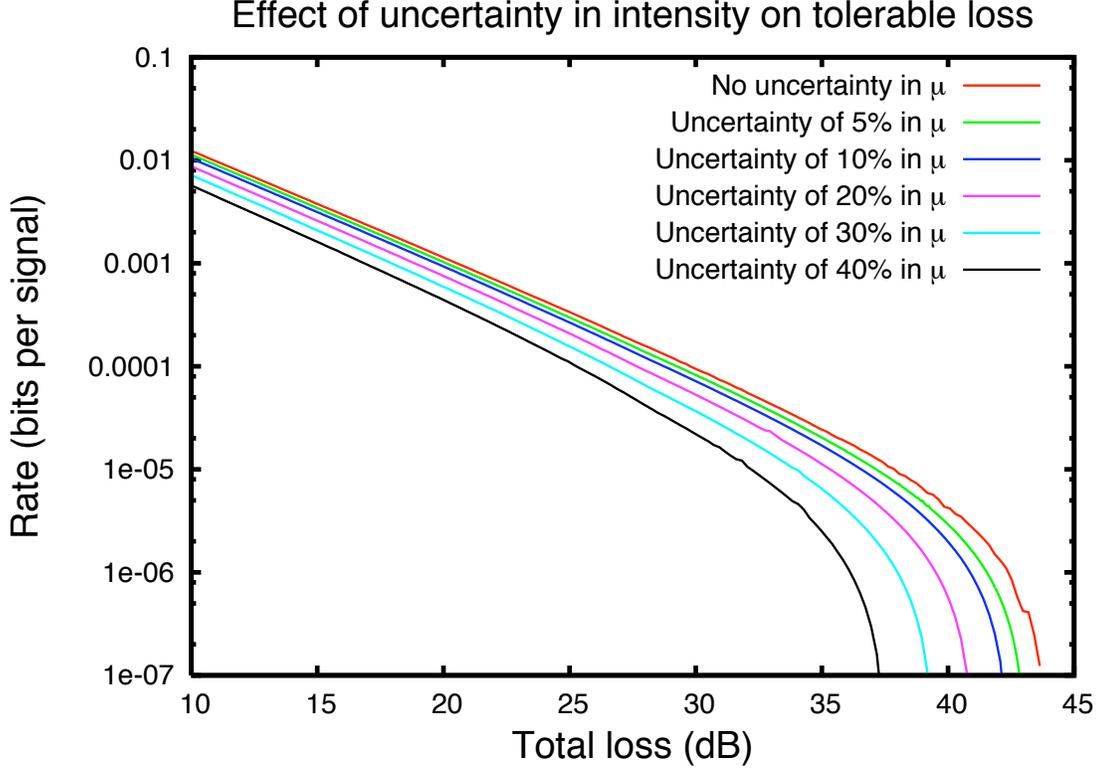}
\caption{(Color online) Plot demonstating how an uncertainty in intensity degrades the secret bit rate.  Fixed system parameters are $N=10^{10}$, $\epsilon = 10^{-7}$, $y_0 = 2 \times 10^{-6}$, and $V = 98\%$.}
\label{graph_uncertainty}
\end{figure}

Here we consider how to incorporate uncertainty into the values of the intensities selected for each level in a decoy state protocol.  In practical terms, the mean photon number of a laser output will never be known exactly, and either the provider or the users of a quantum key distribution system will have to determine how well-calibrated and stable the intensities are.  We model here uncertainty for the low and high intensity levels but do not consider perturbations of the vacuum level.  We first simulate Alice sending the signals through the channel and Bob receiving them with the true intensity values.  Then, for each value $U$ of the uncertainty, we minimize according to the program in Eq. \ref{linearEQ} four separate times using a value of $\mu_j = (1 \pm U) \mu_j$, and taking the worst case of the four results.  We previously ran simulations by modifying both $\mu$ values by a fine grain across this range of uncertainty, and we always found the worst case occurred at the endpoints of the ranges.  Intuitively, this can be seen because only one side of any single inequality from (\ref{linearEQ}) may be tight. Using a non-extreme value for $\mu$ would simply restrict the tight side and thus increase the minimum value. On the other hand, relaxing the loose side of the inequality will not change the minimum value. See Fig.\ \ref{graph_uncertainty} for details on how uncertainty in intensities impacts the secret key rate.  In short, a positive rate exists even with relatively large uncertainties, except when the channel loss is close to the maximum tolerated.

\section{Partially distinguishable levels}

The validity of decoy states and of equations (\ref{linearEQ}) and (\ref{bmin}) is contingent on Eve being unable to guess which signal level Alice sent other than the expected photon number statistics.  If she could distinguish the levels further (due to information leaking into another signal property, such as polarization), then the variables $\{y_{k}\}$ could be modified by Eve to be different values for each level.  We must then expand variables $\{y_{k}\}$ to $\{y_{j,k}\}$ where $j$ is a label for each distinguishable level. The linear program in (\ref{linearEQ}) becomes
\begin{eqnarray}
P_j^{S} &=& e^{-\mu_j} \min \mu_j \overline{y}_1\nonumber \\
P_j^{D} &=& e^{-\mu_j} \min \mu_j \overline{y}_0 \\
\text{subject to:}& \nonumber \\
Y^{-}_{j} &\leq& e^{-\mu_j} \sum_{k=0}^{\infty} \overline{y}_{j,k} \frac{\mu_j^k}{k!} \leq Y^{+}_{j} \ \ \ \forall j \nonumber \\
0 &\leq& \{\overline{y}_{j,k}\} \leq 1\ \ \ \forall j,k. \nonumber
\end{eqnarray}
Now the decoy states contribute no constraints to the optimization. In this case of completely distinguishable decoy states, the protocol does no better than without decoy states. But one could imagine a situation where Eve could only probabilistically distinguish among the decoy states. For example, in a system where the differently encoded bits are created by four different lasers with fixed optics one might want to add a decoy level with a single additional laser. This low intensity laser would need to produce a maximally mixed state (in polarization or phase) to match the average of the four BB84 states.  For signals generated with either zero or one photons, the high and the low intensity states are indistinguishable.  When more photons are generated, these two sources are partially distinguishable. In this example, when Eve cannot distinguish between the states, then a common set of $\{y_k\}$'s are associated with the signal state.  Let us define the probability of indistinguishability as $Q_{j,k}$. This creates additional constraints and the linear program becomes

\begin{figure}[tp]
\includegraphics[width=\columnwidth]{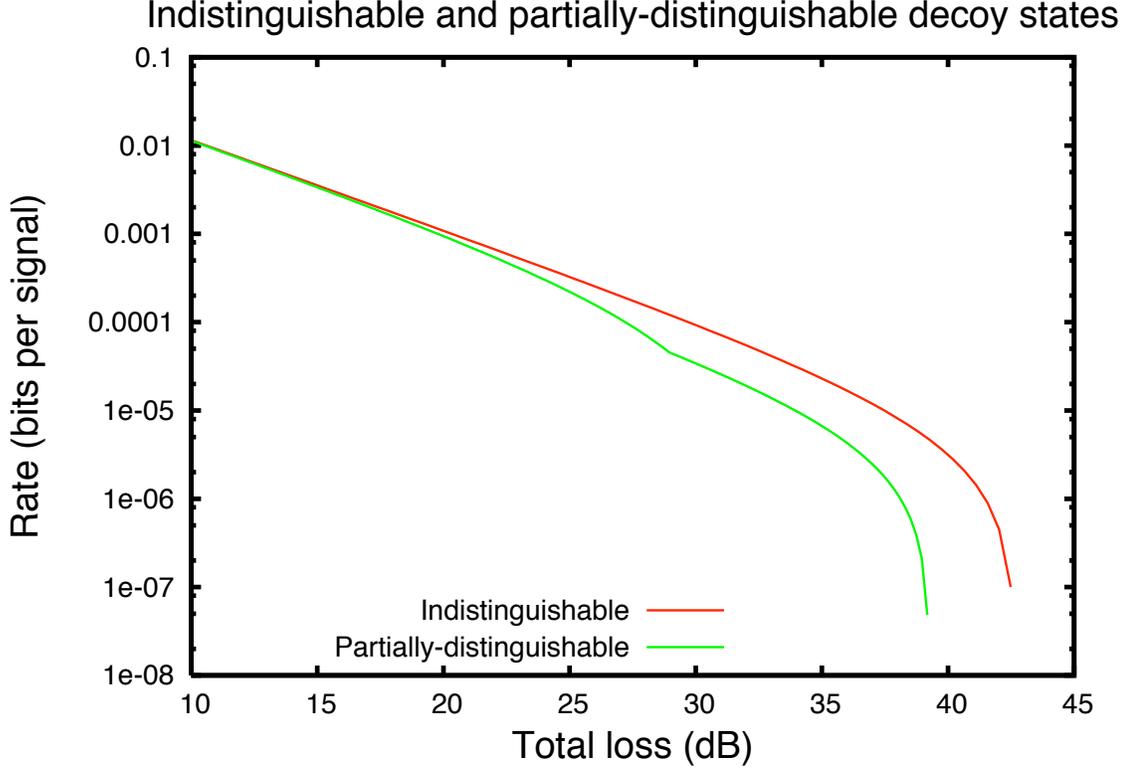}
\caption{(Color online) Semi-log plot of the rate vs.\ the channel loss comparing the partially distinguishable and the undistinguishable decoy state protocols.  Fixed system parameters are $N=10^{10}$, $\epsilon = 10^{-7}$, $y_0 = 2 \times 10^{-6}$, and $V = 98\%$.}
\label{graph_free1}
\end{figure}

\begin{eqnarray}
P_j^{S} &=& e^{-\mu_j} \min \mu_j \overline{y}_1\nonumber \\
P_j^{D} &=& e^{-\mu_j} \min \mu_j \overline{y}_0 \\
\text{subject to:}& \nonumber \\
Y^{-}_{j} &\leq& e^{-\mu_j} \sum_{k=0}^{\infty} \overline{y}_{j,k} \frac{\mu_j^k}{k!} \leq Y^{+}_{j} \ \ \ \forall j \nonumber \\
y_k &\geq& Q_{j,k} \ \overline{y}_{j,k} \ \ \ \forall j,k.\nonumber \\
(1 - y_k) &\geq& Q_{j,k} \ (1 -  \overline{y}_{j,k}) \ \ \ \forall j,k.\nonumber \\
0 &\leq& \{\overline{y}_{j,k}\} \leq 1\ \ \ \forall j,k. \nonumber
\end{eqnarray}

\noindent For the example above, $Q_{j,k} = 1$ when $j$ is the high intensity signal state or the vacuum state. When $j$ is the low signal intensity state $Q_{j,k}$ becomes:

\begin{eqnarray}
Q_{1,k} =
\left\lbrace
\begin{array}{l}
1 \quad  \quad \text{if } k \leq 1 \\
\frac{3}{4} \quad \quad \text{if } k = 2 \\
2^{k-2} \quad \text{if } k \geq 3.
\end{array}
\right.
\end{eqnarray}

Since the low state isn't a key bit carrying signal, it can't have an error rate. This means that the quadratically-constrained program (Eq.\ \ref{bmin}) cannot be used. An upper bound on $b_1$ can be found that is still lower than the worst-case situation. The dark count error rate should be $\frac{1}{2}$ with statistical fluctuations. By putting a lower bound on $y_0$ with the linear program, one can put a lower bound with probability ($1-\epsilon$) that so many errors are due to dark counts. That reduces the number of errors that could be due to single-photon signals. Fig.\ \ref{graph_free1} shows how these limitations effect the rate compared to the standard three-level protocols for the example $Q$'s.
 
\section{Discussion}

These simulations shed some light on what system parameters are most important to improve rates and maximum distances.  We find that on the order of $10^{5}$ received detections are sufficient for a positive secret bit rate with our finite statistics approach, provided that noise is sufficiently low.  The intensities for 3-level decoy state protocols are generally optimized with $\mu_0 = 0$ (vacuum) and $\mu_1 \sim 0.1$, while the optimal value for the key-generating signal intensity is dominated by the noise, with a rough fit of $\mu_2 \sim e^{-15({\rm BER})}$.  Modifying the total loss by an order of magnitude results in only minor modifications to these values.  Increasing the security parameter by many orders of magnitude also leads to very minor effects, so there is little cost to boosting confidence levels (provided that there are sufficient statistics collected to bound the single photons and their bit error rate).  As expected, dark counts are the limiting factor for maximizing the distances (or tolerable loss) for a positive secret bit rate.  It is worth noting, however, that an order of magnitude suppression in dark and background count rates generally adds less than 10 dB to the tolerable loss, unless the session length is also increased to maintain count statistics.  Various imperfections can also be incorporated into the decoy state analysis, such as not knowing the intensities exactly or potentially leaking partial information about the levels to an adversary, and all of the bounds described in this work can be implemented in software to run in real-time.

\appendix
\section{Why bounds derived for stationary channels are valid for non-stationary channels}
\label{stationaryArgument}

The bounds created by the linearly- and quadratically-constrained programs were obtained with the assumption that Eve provides a stationary channel characterized by the parameters $\{y_k, b_k\}$. This assumption is not necessary.  We will show that the bounds are valid in the situation that for each signal Alice sends, Eve chooses one of $L$ channels to apply (for arbitrary $L$). The $L$ channels are parameterized by $\{y_{k,l}, b_{k,l}\}$ and Eve applies them with frequencies $f_l$, with $\sum_{l = 1}^L f_l = 1$.

\subsection{Yield bounds}
In the stationary case, the total number of single-photon signals (SPS) is $(p_1\ N\ y_1)$ and the total number of clicks is $(N\ \sum_k p_k y_k)$. The inequality for each signal intensity appearing in the linear program becomes
\begin{eqnarray}
Y^- \leq \sum_k p_k \ \overline{y}_k \leq Y^+ ,
\end{eqnarray}
where $Y^{\pm}$ are the bounds on the yield $\frac{C}{N}$, $\overline{y}_k$ are the variables over which the program is optimized, and $p_k$ is the probability that a signal contains $k$-photons. Let the functions that calculate the yield bounds be $g_-(s,t,\epsilon)$ and $g_+(s,t, \epsilon)$.  Specifically, they provide lower and upper bounds on the success probability of a binomial trial, given that for $t$ trials the observed frequency was $\frac{s}{t}$. The statement that the true success probability lies between both bounds has a confidence of at least $(1-\epsilon)^2$.

Now, when Eve uses the $L$ channels the true value of $y_1$ becomes $\sum_l f_l \ y_{1,l}$, the total number of the SPS becomes $\left(N \ p_1 \ \left(\sum_l f_l \ y_{1,l}\right)\right)$ and the total number of clicks become $C = N\ \sum_k\left(\sum_{l=0}^L f_l \ y_{1,l} \right)$. The inequalities are:
\begin{eqnarray}
\sum_k p_k \ \overline{y}_k 
&\geq&
g_-\left(N \ \sum_k\left( \sum_{l=0}^L f_l \ y_{1,l} \right), N, \epsilon\right) 
\nonumber \\
\sum_k p_k \ \overline{y}_k 
&\leq& 
g_+\left( N \ \sum_k\left( \sum_{l=0}^L f_l \ y_{1,l} \right), N, \epsilon\right)
\end{eqnarray}

Defining new variables $\overline{y}_{k,l}$, $g_{-,l}$, $g_{+,l}$ such that  $\sum_l f_l\overline{y}_{k,l} =  \overline{y}_k$, $\sum_l f_l\ g_{-,l} = g_-(\cdot)$, and $\sum_l f_l \ g_{+,l} = g_+(\cdot)$,  we can create an equivalent linear program
\begin{eqnarray}
P &=& \min~\sum_{l=1}^L \overline{y}_{k,l} \\
\text{subject to:} \nonumber \\
\sum_l g_{-,l}
&\leq& 
\sum_k \sum_l f_l \ p_k \ \overline{y}_{k,l}  
\nonumber \\
\sum_l g_{+,l}
&\geq& 
\sum_k \sum_l f_l \ p_k \ \overline{y}_{k,l}
\nonumber \\
\overline{y}_{k,l} 
&\geq&
0
\nonumber \\
\overline{y}_{k,l} 
&\leq&
1
\label{original}
\end{eqnarray}

Now we can create $L$ fictional linear programs $\min \overline{y}_{k,l}$ subject to:
\begin{eqnarray}
&\left( g_{-,l}\right) \leq& \sum_k\left( f_l \ p_k \ \overline{y}_{k,l} \right) \leq \left( g_{+,l}\right). \nonumber \\
&&0 \leq \{ \overline{y}_{k,l} \}\leq 1 \label{fictional}.
\end{eqnarray}
Since the inequalities in (\ref{fictional}) are stronger that the inequality created by their sum found in the original program, the polytope of allowable solutions created by the intersection of the polytope (over all variables) for each program is contained in the polytope of allowable solutions to (\ref{original}). This means that the original program provides a lower bound on the sum of the solutions to the fictional programs.
\begin{eqnarray}
\overline{y}_1^* \leq \sum_l \overline{y}_{1,l}^* \label{bound1}.
\end{eqnarray}
The fictional programs provide bounds for the $L$ stationary channels. So we have
\begin{eqnarray}
 \overline{y}_{1,l}^* \leq f_l \ y_{1,l}^* \label{bound2}.
\end{eqnarray}
Putting (\ref{bound1}) and (\ref{bound2}) together completes the argument
 \begin{eqnarray}
\overline{y}_1^* \leq \sum_l f_l \ y_{1,l}^*  \leq \sum_l f_l \ y_{1,l} = y_1.
\end{eqnarray}

\subsection{Disturbance bounds}

In the stationary case, the true value of $b_1$ is simply $b_1$. That can be rewritten to $\frac{b_1 \ y_1}{y_1}$ for positive $y_1$. Defining $c_k = b_k \ y_k$ as the total probability that a $k$-photon signal sent from Alice will result in an erroneous click at Bob, $b_1$ becomes $\frac{c_1}{y_1}$. The total number of errors is $(N\ \sum_k p_k c_k)$. The inequalities of the program are

\begin{eqnarray}
Y^- \leq \sum_k p_k \ \overline{y}_k \leq Y^+  \nonumber \\
B^- \leq \sum_k p_k \ \overline{b}_k \ \overline{y}_k \leq B^+ \label{ber_basic}
\end{eqnarray}

\noindent where $B^{\pm}$ are bounds on each signal intensity's bit error rate. 

Now, when Eve uses the $L$ channels, the true value of $b_1$ is still $\frac{c_1}{y_1}$, but now $c_1 = \sum_l f_l \ c_{1,l}$. An upper bound on $b_1$ is

\begin{eqnarray}
b_1 = \frac{\sum_l f_l\ c_{1,l}}{y_1} \leq \frac{\max\{\overline{c}_1\}}{\min\{\overline{y}_1\}} \leq \frac{\max\{\sum_l f_l\ \overline{c}_{1,l}\}}{\overline{y}_1^*}.
\end{eqnarray}

\noindent This means that we can find an upper bound on $b_1$, if we can find an upper bound on $c_1$. 

We can transform (\ref{ber_basic}) into an equivalent quadratically-constrained program with the following inequalities

\begin{eqnarray}
\sum_k p_k \ \overline{y}_k 
& \geq &
g_-\left(\left( N \ \sum_k \sum_l f_l \ p_k \ y_{k,l} \right), N, \epsilon\right) 
\nonumber \\
\sum_k p_k \ \overline{y}_k 
& \leq &
g_+\left(\left(N \ \sum_k \sum_l f_l \ p_k \ y_{k,l} \right), N, \epsilon\right)  
\nonumber \\
\sum_k p_k \ \overline{c}_k 
& \geq &
g_-\left(\left(N \ \sum_k \sum_l  f_l \ p_k \ c_{k,l} \right), N, \epsilon\right) 
\nonumber \\
\sum_k p_k \ \overline{c}_k 
& \leq & 
g_+\left(\left(N \ \sum_k \sum_l f_l \ p_k \ c_{k,l} \right), N, \epsilon\right) 
\nonumber \\
\overline{c}_k 
&=& 
\overline{b}_k \ \overline{y}_k \label{ber}.
\end{eqnarray}

The objective function is still $b_1$, but note that any assignment of variables that maximizes $b_1$ will also maximize $c_1$ since $b_1$ only appears next to $y_1$ in the inequalities. This can be proved by contradiction. Assume an assignment of variables that maximizes $b_1$. This implies that at least one constraint that involves $b_1$ is tight. $c_1$ can only be increased by either increasing $b_1$ or $y_1$. If $b_1$ can be increased, then it wasn't maximized. If $y_1$ can be increased, all inequalities that involve $y_1$ are slack. Since $b_1$ is only in inequalities that involve $y_1$ this is a contradiction. Making a trade-off such as first decreasing one variable in order to increase the other can't work either, since any such trade-off keeps $c_1$ constant.

Now, we can play the same trick as before. Again, the intersection of the allowable solutions to the $L$ fictional programs is contained in the region (no longer necessarily a polytope) allowed by (\ref{ber}). So the maximum of the original program is an upper bound on the sum of the fictional maximums. $c_1^* \geq \sum_l c_{1,l}^*$. Solving the fictional programs yields

\begin{eqnarray}
 \overline{c}_{1,l}^{*} &\geq& f_l \ c_{1,l} \label{ber_bound}.
\end{eqnarray}

Combining with the above means that the original program provides an upper bound on $c_1$ and thus an upper bound on $b_1$.

\bibliography{NumericalAnalysisDecoyQKD}

\end{document}